\documentclass[
aps,
twocolumn,prl,
showpacs,preprintnumbers,nofootinbib,
amsmath,amssymb,
aps,
superscriptaddress]{revtex4-1}
\usepackage[breaklinks=true]{hyperref}
\usepackage{breakcites}
\usepackage{float}
\usepackage{graphicx}
\usepackage{dcolumn}
\usepackage{bm}
\usepackage{hyperref}
\usepackage{ulem} 
\usepackage{breakcites}

\newcommand{\lm}{{\ell,m}}

\newcommand{\be}{\begin{equation}}
\newcommand{\ee}{\end{equation}}
\newcommand{\bi}{\begin{itemize}}
\newcommand{\ei}{\end{itemize}}
\newcommand{\bea}{\begin{eqnarray}}
\newcommand{\eea}{\end{eqnarray}}

\usepackage{color}

\begin{document}

\title{Tracking black hole kicks from gravitational wave observations}

\affiliation{Center for Relativistic Astrophysics and School of Physics, Georgia Institute of Technology, Atlanta, GA 30332}
\affiliation{Monash Centre for Astrophysics, School of Physics and Astronomy, Monash University, VIC 3800, Australia}
\affiliation{OzGrav: The ARC Centre of Excellence for Gravitational-Wave Discovery, Clayton, VIC 3800, Australia}

\author{Juan Calder\'on~Bustillo$^{1,2,3}$}\noaffiliation
\author{James A. Clark$^\text{1}$}\noaffiliation
\author{Pablo Laguna$^\text{1}$}\noaffiliation
\author{Deirdre Shoemaker $^\text{1}$}\noaffiliation

\preprint{LIGO-P1800179}

\begin{abstract}

Coalescing binary black holes emit anisotropic gravitational radiation. This causes a net emission of linear momentum that produces a gradual acceleration of the source. As a result, the final remnant black hole acquires a characteristic velocity known as recoil velocity or gravitational kick. The symmetries of gravitational wave emission are reflected in the interactions of the gravitational wave modes emitted by the binary. In this work we make use of the rich information encoded in the higher-order modes of the gravitational wave emission to infer the component of the kick along the line-of-sight (or \textit{radial kick}). We do this by performing parameter inference on simulated signals given by numerical relativity waveforms for non-spinning binaries using numerical relativity templates of  aligned-spin (non-precessing) binary black holes. We find that for suitable sources, namely those with mass ratio $q\geq 2$ and total mass $M \sim 100M_\odot$, and for modest radial kicks of $120km/s$,  the $90\%$ credible intervals of our posterior probability distributions can exclude a zero kick at a signal-to-noise ratio of $15$; using a single Advanced LIGO detector working at its early sensitivity.  The measurement of a non-zero radial kick component would provide the first observational signature of net transport of linear momentum by gravitational waves away from their source. 

\end{abstract}
\maketitle


\paragraph*{\textbf{Introduction}}

The detection of gravitational waves (GWs) from coalescing binary black holes (BBHs) by the Advanced LIGO \cite{TheLIGOScientific:2014jea} and Virgo \cite{TheVirgo:2014hva} detectors has opened the long anticipated era of GW astronomy
\cite{Abbott:2016blz,Abbott:2016nmj,PhysRevLett.118.221101,Abbott:2017oio,Abbott:2017gyy}. With LIGO soon entering its third observation run with improved sensitivity \cite{Aasi:2013wya,advLIGOcurves} it is expected that BBH detections will not only become more frequent but also louder. This will  improve studies on the astrophysical distribution and origin of  BBHs~\cite{TheLIGOScientific:2016htt,Farr:2017uvj}  and enhance tests of General Relativity (GR) \cite{TheLIGOScientific:2016src,PhysRevX.6.041015,PhysRevLett.118.221101}. GR predicts that gravitational waves carry linear momentum away  from the source \cite{Misner:1974qy,Bekenstein:1973zz,Ruiz:2007yx}; however no measurement of this feature has yet been performed.  In this letter we propose and demonstrate a simple method to prove the existence of a net emission of gravitational wave linear momentum emitted by BBHs via the measurement of the recoil velocity (or \textit{kick}) of the remnant black hole (BH) relative to the observer\cite{Gonzalez:2006md,Brugmann:2007zj,Lousto:2007db,Lousto:2011kp,Healy:2008js}.\\

\noindent The two ``$+$'' and ``$\times$'' polarizations of a GW emitted by a BBH coalescing at a time $t_c$ can be written as a superposition of GW modes, $h_\lm$, weighted by spin $-2$ spherical harmonics  $Y^{-2}_\lm$ as 
\begin{equation}
\begin{aligned}
h_+-ih_{\times} =\sum_{\ell\geq 2}\sum_{m=-\ell}^{m=\ell}Y^{-2}_{\ell,m}(\iota,\varphi)h_{\ell,m}(\Xi;t-t_c),
\end{aligned}
\label{gwmodes}
\end{equation}
Here, $\Xi$ denotes the masses $m_i$ and dimensionless spins $\vec\chi_i$ of the individual components of the BBH and $(\iota,\varphi)$ are spherical coordinates describing the location of the observer around it (or conversely, the orientation of the binary). For non-precessing binaries, and during most of the coalescence process, the above sum is vastly dominated by the $(2,\pm 2)$ modes. The others, known as higher modes (HMs), become strong only during the final stages of the process, increasing their impact as the mass ratio $q=m_1/m_2 \geq 1$ grows \cite{Bustillo:2015qty,Bustillo:2015ova}. In general, GW modes interact in an anisotropic fashion, leading to a net emission of GW linear momentum which increases during the merger stage of the coalescence \cite{Gonzalez:2006md}, when both the net GW emission and the HMs become stronger.  As a consequence, the final BH acquires a characteristic final velocity known as BH kick $\vec K$. Its magnitude $|K|$ depends on the mass ratio and the individual spins of the objects and can reach values of $\sim 10^3km/s$ in the most extreme cases \cite{Bekenstein:1973zz,Brugmann:2007zj,Sperhake:2010uv, Healy:2008js,Campanelli:2007ew,Campanelli:2007cga}, enough to make the final BBH escape its host galaxy \cite{Merritt:2004xa,Gerosa:2014gja}. 

\begin{figure}
\includegraphics[width=0.475\textwidth]{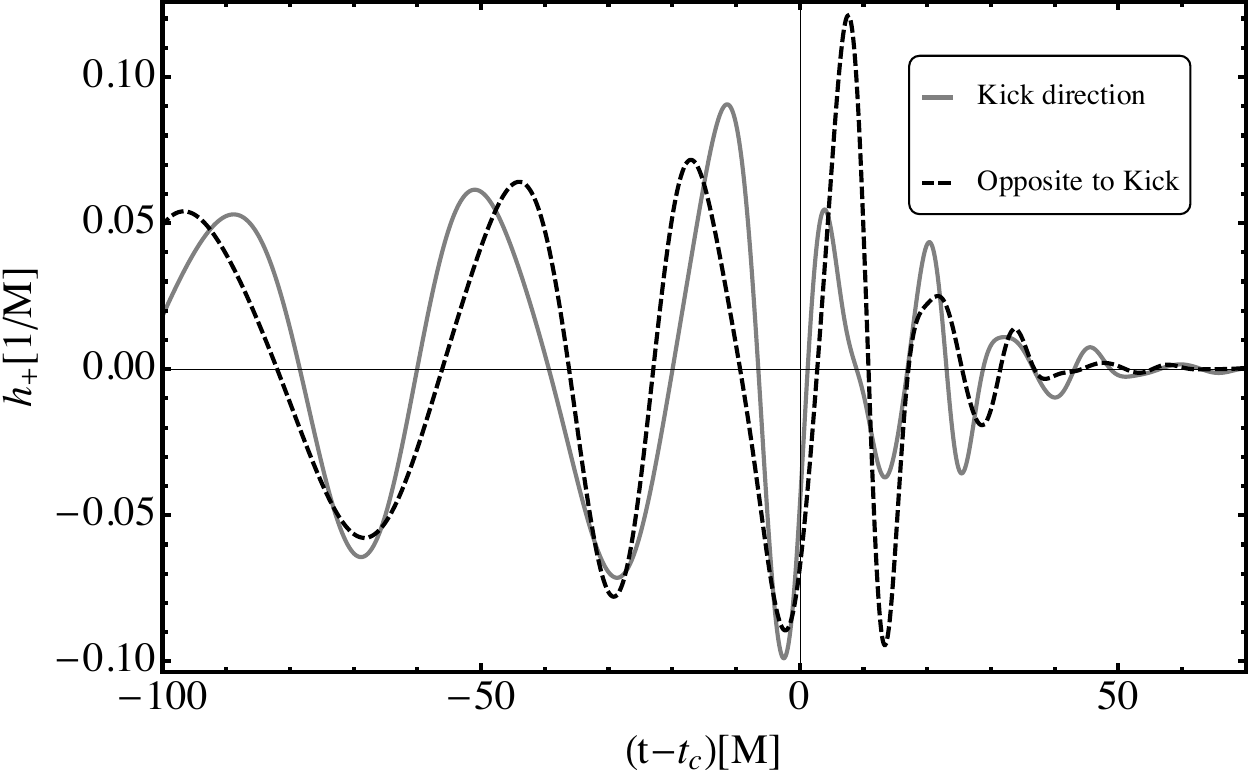}
\caption{\textbf{Impact of azimuthal angle}: Last cycles of a GW signal emitted by a $q=3$ non precessing binary in the direction of its final kick (solid) with $(\iota,\varphi)=(\pi/2,\varphi_{K})$ and opposite to it (dashed) with $(\iota,\varphi)=(\pi/2,\varphi_{K}+\pi)$. Strain and time are expressed in numerical relativity units and $t_c$ coincides with the amplitude peak of the $(2,2)$ mode. The observed final velocities are $K_r = \pm 163$km/s. See section II for a detailed definition of the angles.
}
\label{fig:strains}
\end{figure}

A method to estimate $K_r$ was proposed in \cite{Gerosa:2016vip}. There, the effect of the kick was modelled as a \textit{progressive Doppler shift on the observed dominant $(2,2)$ mode of the GW emission}. However, since the kick is caused by the asymmetric interaction of GW modes, any imprint of $K_r$ in the GW signal should be included by adding all modes together and no artificial imparting of a Doppler shift should be needed to account for the kick\footnote{Note that by definition, the $(2,2)$ and all other modes look the same in all directions for a given source (see Eq.(1)).}. Furthermore, because different observers around the source observe a very different interaction between the multiple GW modes, the difference between signals corresponding to different $K_r$ (or observed at different locations) is more dramatic than a simple shift in the observed frequencies.  For instance, for the case shown in Fig.\ref{fig:strains}, the observer located in the direction of the final kick measures a weaker signal peak than the one located opposite to it, which can not be accounted by means of Doppler shifts (see Fig.1 in \cite{Gerosa:2016vip})\footnote{A study of the morphology of these signals and the physics behind it is beyond the scope of this work. However, an article addressing these details is already in preparation \cite{Requiem}.}. \\

In this letter we will exploit these features to infer the location of the observer with respect to the final kick. Next, we will combine this with estimates of $|K|(\Xi)$ obtained via the estimation of the BBH intrinsic parameters $\Xi$ to estimate $K_r$.  
 
 \paragraph*{\textbf{Black hole kick frame}}

For non-precessing binaries the direction of the angular momentum $\vec L$ is conserved, making it natural to define the polar (or inclination) angle $\iota$ from eqn.~(\ref{gwmodes})  as the angle between $\vec L$ and the  line-of-sight. This way, $\iota=0$ is parallel to $\vec L$ (source face-on to the observer) and $\iota=\pi/2$ defines the orbital plane (source edge-on). While it is common practice to report the inferred value of $\iota$ for gravitational wave observations \cite{TheLIGOScientific:2016wfe}, the azimuthal angle $\varphi$ is generally regarded as a nuisance parameter due to its lack of physical significance in most analyses, which only consider the dominant $(2,2)$ mode. In this case, varying $\varphi$ only leads to global phase shift in the observed signal, which makes the GW radiation isotropic and difficults the definition of a physical phenomenon that defines its origin. However, when including HMs, the waveform morphology depends on $\varphi$ on a more complex way.\\ 
\noindent For aligned-spin binaries the kick $\vec K$ is contained in the orbital plane, defining a preferred direction in it given by 
  $(\iota=\pi/2,\varphi=\varphi_K)$. This allows us to define a $\textit{kick frame of reference}$ with angular coordinates $(\iota,\bar\varphi= \varphi - \varphi_K)$. This way, the azimuth $\bar\varphi$ measures the angle between the final kick and the projection of the line-of-sight onto the orbital plane. The component of the kick onto the line-of-sight (or \textit{radial kick}) may then be written as $K_r=K(\Xi)\cos{\alpha}=K(\Xi)\cos{\bar\varphi}\cos(\iota-\pi/2)$.\\  

\begin{figure}
\includegraphics[width=0.5\textwidth]{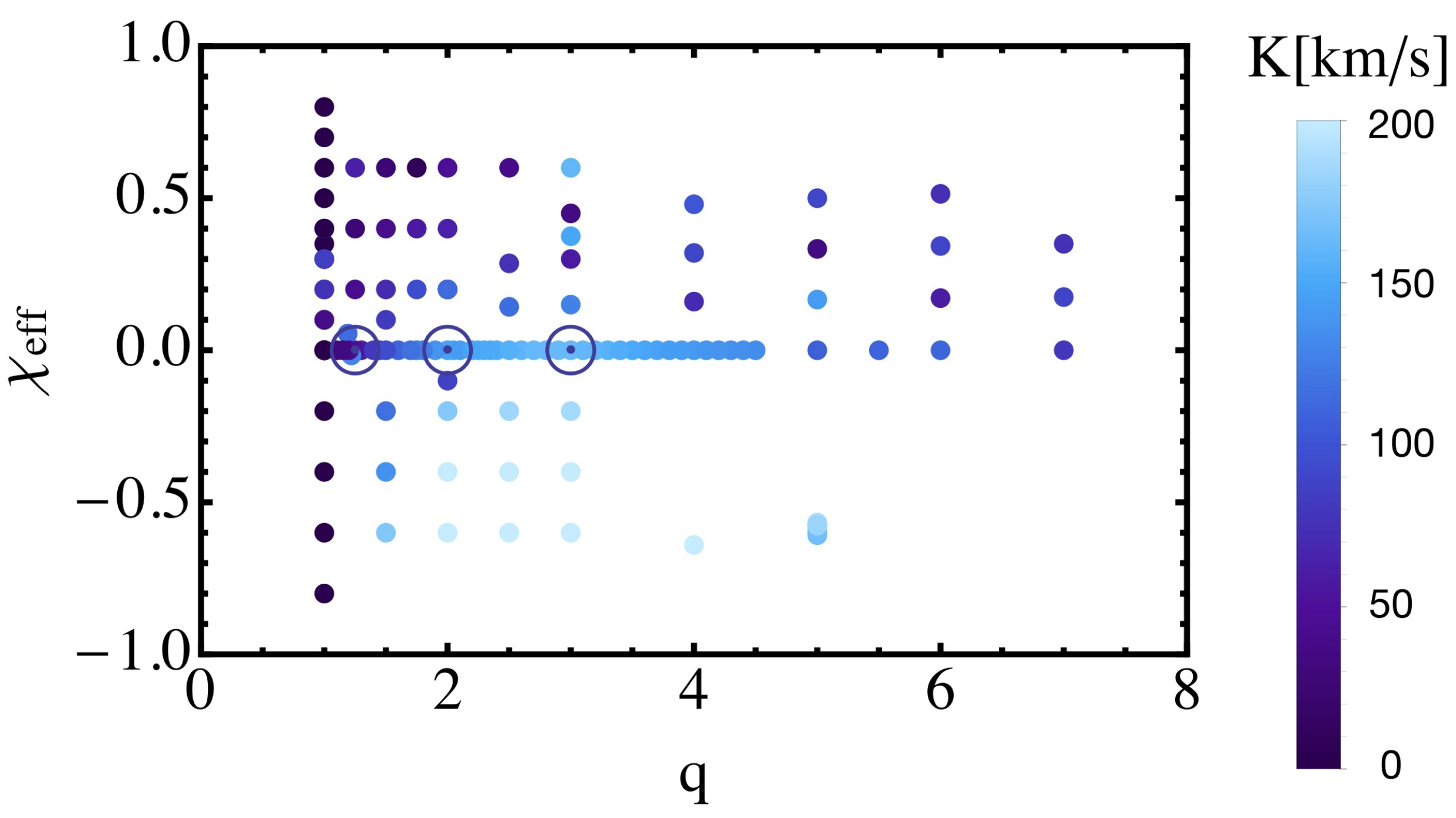}
\caption{\textbf{Bank of non-precessing numerical simulations}: Set of simulations used as templates  and injections (black circles) in this study, represented by their $q$ and effective spin $\chi_{eff}=\frac{m_1\chi_{1,z}+m_2\chi_{2,z}}{m_1+m_2}$. Here, $\chi_{i,z}$ denotes the projection each BH spin along the direction of the orbital angular momentum.
}
\label{fig:catalogue}
\end{figure}

\begin{figure*}
\includegraphics[width=0.46\textwidth]{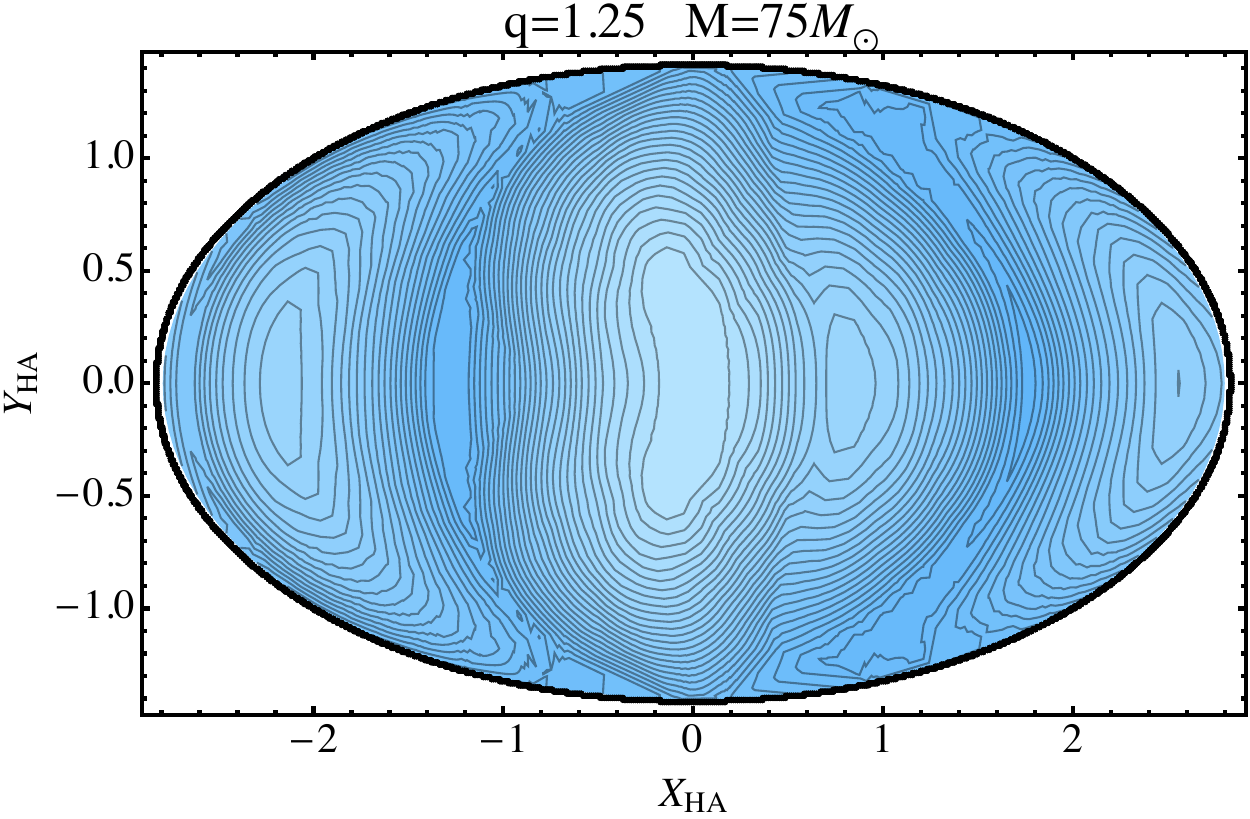}
\includegraphics[width=0.46\textwidth]{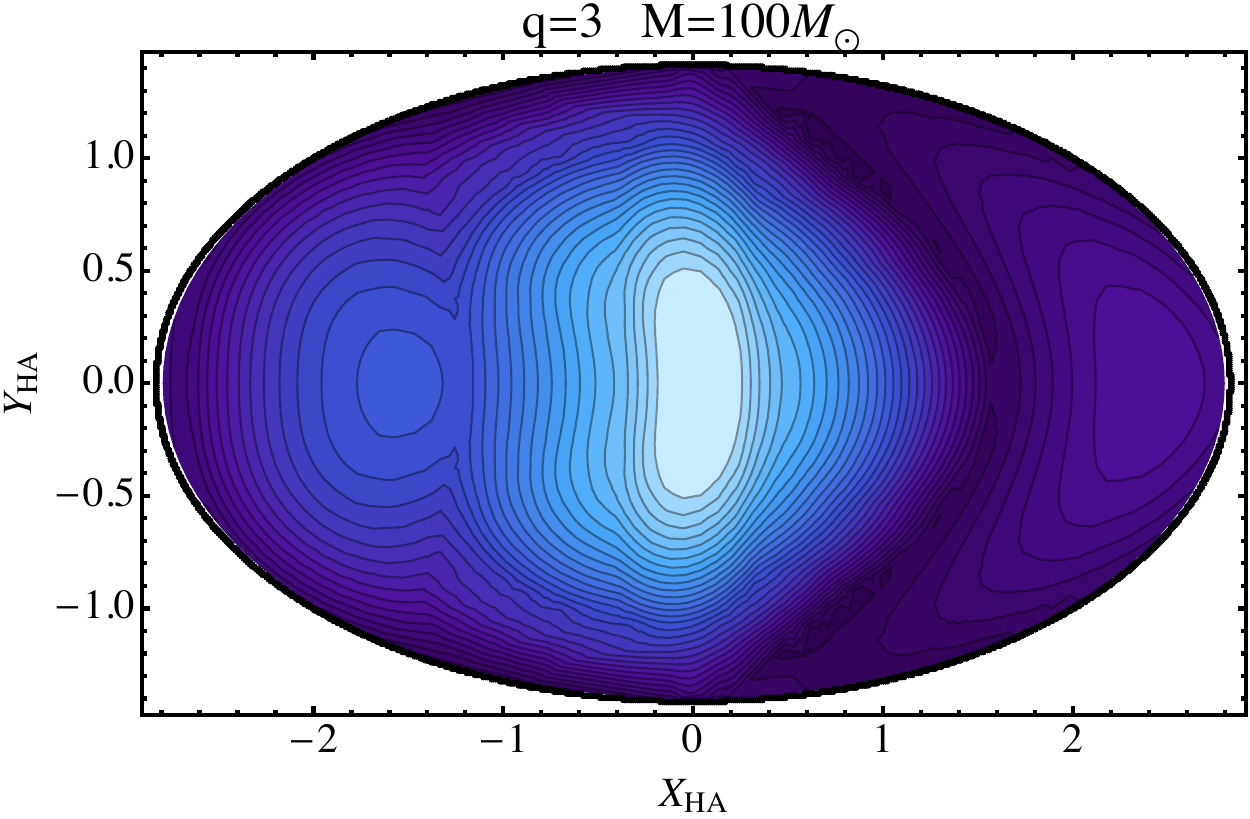}
\includegraphics[width=0.065\textwidth]{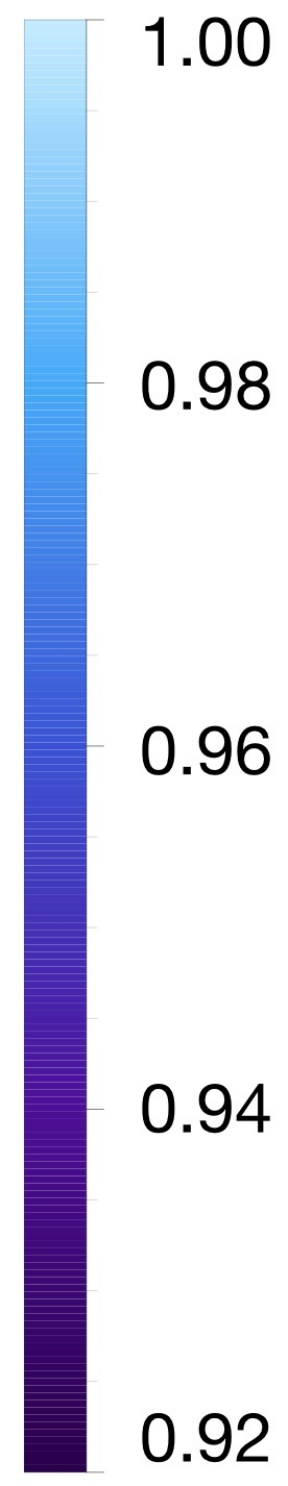}
\caption{\textbf{Resolvability of orientation}: Overlap, maximised over masses and spins, between two fiducial 
signals emitted in the direction of the final kick of two binaries and our NR template bank as a function of the 
orientation of the templates, expressed in Hammer-Aitoff coordinates. 
$Y_{HA}=0$ denotes edge-on orientations (or the orbital plane) while 
$Y_{HA}=\pm 1$ denotes face-on/off orientations. $X_{HA}$ is defined so that 
$X_{HA}=0$ corresponds to $\bar\varphi=0$. Hence, $(X_{HA},Y_{HA})=(0,0)$ denotes the direction of the final kick. The fiducial signals correspond to a binary with parameters 
consistent with GW150914 (left panel) and a $q=3$ binary with a total mass of 
$M=100M_\odot$. The stronger higher mode content of the $q=3$ binary 
leads to a stronger dependence of the signal on the orientation of the binary, 
and a rapid degradation of the overlap for orientations different than that of 
the source.}
\label{fig:contours}
\end{figure*}

\paragraph*{\textbf{Analysis}}
Our goal is to determine the precision with which we can measure $K_r$, 
given time-domain data $d(t)=n(t)+h(t;\vec{\lambda})$, where $n(t)$ is 
environmental and instrumental noise, and $h(t;\vec{\lambda})$ is the waveform 
of a potential GW signal with source parameters $\vec{\lambda}=\{\Xi;\iota,\bar\varphi;t_c ...\}$. Our figure of merit will be the confidence intervals of the marginalized posterior probability of $K_r(\vec\lambda)$. To obtain this, we first compute the posterior 
probability density function (PDF) for $\vec\lambda$, denoted by $p(\vec{\lambda}|d) \sim 
\pi(\vec{\lambda}){\cal L}(d|\vec{\lambda})$.  With this, we can then compute the marginal 
posterior on $K_r$ as  $p(K_r|d)=\int_{\Lambda(K_r)} p(\vec{\lambda}|d) 
 d\vec\lambda$, with 
$\Lambda(K_r)=\{\vec\lambda : K_r(\vec\lambda)=K_r\}$. Here, $\pi(\vec{\lambda})$ is the 
prior PDF on the source parameters and ${\cal L}(d|\vec{\lambda})$ is 
likelihood of our observations  $d$, given source parameters $\vec\lambda$.  
Simplifying our notation, the likelihood is given by ${{\cal 
L}}(\vec\lambda)=e^{-\frac{1}{2}(d-h(\vec\lambda)|d-h(\vec\lambda))}$  
\cite{Cutler:1994ys} where $(a|b)$ denotes the noise weighted inner product $4  
{\cal R}\int_{f_0}^{f_h} \tilde{a}(f)\tilde{b}^{*}(f)/S_n(f) df$ and $S_n(f)$  
is the one sided spectral density of the detector noise \cite{MatchedFilter}. In this study we consider the case of Advanced LIGO working at its predicted early sensitivity \cite{Aasi:2013wya} with a lower frequency cutoff $f_0=24$Hz. \\
The waveforms $h(\vec\lambda)$ used in this study are obtained from the GeorgiaTech 
catalog of numerical relativity (NR)
simulations~\cite{Jani:2016wkt,Bustillo:2016gid} \footnote{We include the most 
dominant modes, namely those with $(\lm)=\{(2,\pm 1),(2,\pm 2),(3,\pm 2),(3,\pm 
3),(4,\pm 4)\}$.}, which cover the parameter space shown in 
Fig.~\ref{fig:catalogue}. 
In these simulations, signals are extracted on a 
fixed sphere that does not follow the center-of-mass of the BBH. Hence, the 
information about the kick is already present in the extracted modes and its effect is correctly implemented in the resulting waveforms. In the evaluation of ${\cal L}(d|\vec{\lambda})$, we 
set the noise term $n(t)=0$, which is equivalent to averaging 
the posterior PDF over many noise realizations~\cite{Nissanke:2009kt} and can 
therefore be used to determine the expected precision for the measurement of 
$K_r$.   
Finally, we characterize the intrinsic loudness of our signals $h$ via 
the optimal signal-to-noise ratio (SNR)  $\rho = (h|h)^{1/2}$ and the 
similarity of two signals $h_1(\vec{\lambda_1})$ and $h_2(\vec{\lambda_2})$  is quantified in terms of their overlap 
$O(h_1|h_2)=(h_1|h_2)/\sqrt{(h_1|h_1)(h_2|h_2)}$ \cite{Apostolatos:1994mx}.

To evaluate the posterior PDF $p(\vec{\lambda}|d)$ we adopt uniform, independent 
priors on the intrinsic source parameters and, for each point shown in Fig.~\ref{fig:catalogue}, we evaluate ${{\cal 
L}}(\vec\lambda)$ over a grid of orientations $(\iota,\bar\varphi)$ and total 
mass $M=m_1+m_2$ with a resolution of $\delta M =0.5M_\odot$, 
$\delta\bar\varphi=0.1$ and $\delta(\cos\iota)=0.1$. Finally, as in similar studies, we maximize ${{\cal 
L}}(\vec\lambda)$ over time of arrival, polarization angle 
and distance \cite{Lange:2017wki}.   
In the sections which follow, we 
report and study credible intervals for $K_r$ for a variety of simulated test signals, setting $d(t) = h_{inj}(\vec\lambda_{inj})$.  Following 
the parlance of GW data analysis, we refer to $h_{inj}$ as \textit{injections} and to $h(\vec\lambda)$ as \textit{templates}.

Unlike analytic approximate models \cite{Khan:2015jqa,Bohe:2016gbl}, NR 
waveforms cover a discrete family of BBH parameters, which limits our sampling 
of ${\cal{L}}(\vec\lambda)$. Because our NR  bank is more dense in the $q<3$ 
region, we confine our study to this part of the parameter space.  We perform 
injections with intrinsic parameters $q=2,3$ and 
$100M_\odot$ and $200M_\odot$. In addition, we consider a source with 
parameters marginally consistent with GW150914 \footnote{The corresponding BBH simulations are labeled as GT0446, GT0453 and 
GT0738 in the public GeorgiaTech waveform catalogue in http://www.einstein.gatech.edu/table .}. 
 \\ %

\paragraph*{\textbf{Results}}In previous work \cite{Gerosa:2016vip}, the distinguishability of non-zero $K_r$ waveforms was assessed by means of their mismatch $(1-O)$ to $K_r=0$ ones. Modelling the effect of the kick via Doppler shifts, $K_r$ of the order of $10^3$km/s lead to mismatches of $10^{-5}$. This means that SNRs of $\rho\sim 1/\sqrt{10^{-5}} \sim 300 $ would be needed to measure such kicks \cite{Lindblom:2008cm}. In contrast, modelling the kick including the HMs causes a stronger imprint in the waveform that facilitates its measurement. Fig.\ref{fig:contours} shows the overlap between two waveforms (or injections) emitted in the direction of the final kick of two different BBHs to \textit{all} our NR bank of templates $h(\Xi;\iota,\bar\varphi)$ as a function of $(\iota,\bar\varphi)$ and maximised over $\Xi$.  In the right panel, the injection has $(q^{inj},M^{inj})=(3,100M_\odot)$ and $K_r^{inj} \sim 160$km/s. The strong HMs produce a strong dependence of the signal on the BBH orientation, leading to a fast degradation of the overlap as $(\iota,\bar\varphi)$ differ from those of the injection, and to a better resolvability of the orientation. In particular, the mismatch between our injections and $K_r=0$ templates can be as low as $1-0.92=0.08$, thus distinguishable at an SNR $\rho \sim 4$. In contrast, since HMs are much weaker for a source like GW150914, overlaps are always greater than $0.985$ making the orientation harder to constrain (left panel). \\

\begin{figure}
\includegraphics[width=0.5\textwidth]{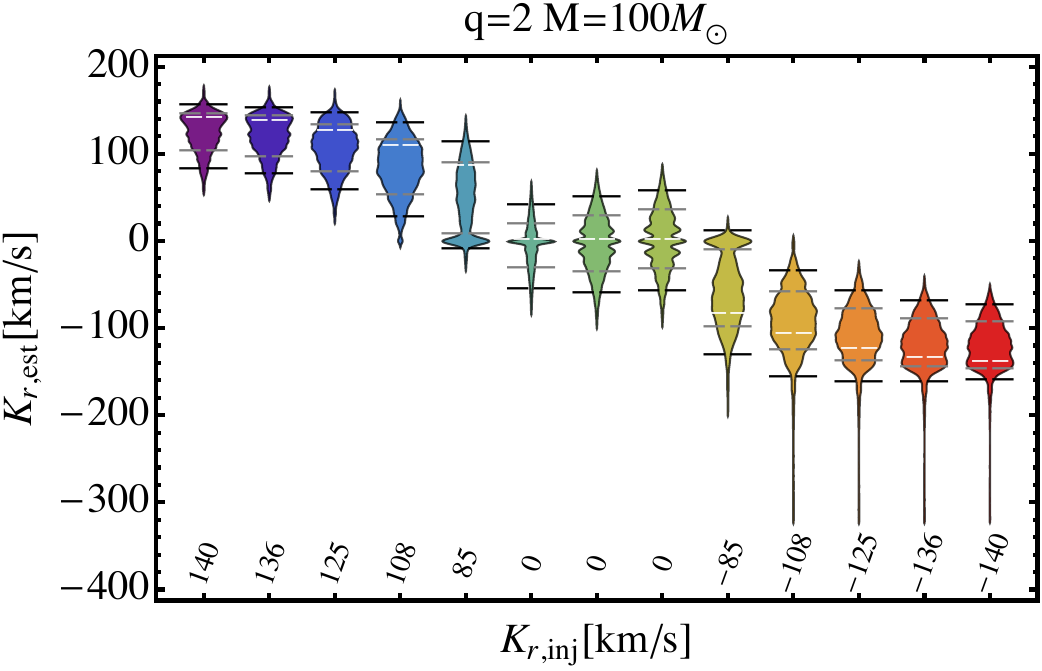}
\caption{\textbf{Estimations of $K_r$ for selected injections}: Posterior distributions for $K_r$ for several injections corresponding to a non-spinning BBH with $q=2$ $M=100M_\odot$ with varying orientations with respect to the observer. We consider an observed SNR $\rho=15$. 
}
\label{fig:violins}
\end{figure}

\noindent Although the mismatch provides a simple method to assess the measurability of a parameter, it is more appropriate to do this using full Bayesian parameter inference. 
Fig.\ref{fig:violins} shows posterior distributions for $K_r$ for signals emitted by a non-spinning binary with $(q,M)=(2,100M_\odot)$, varying orientation and optimal SNR $\rho=15$.\footnote{We obtain the continuous distributions shown in the figure applying a Gaussian kernel density estimator \cite{MathematicaKDE} on our discrete NR sample grid.} The real value $K_{r,inj}$ of the signal (white bars) is always well within the $68\%$ and $90\%$ credible regions of our posterior distributions, delimited respectively by grey and black bars.\footnote{These are respectively defined as the intervals delimited by the  16th-84rd and 5th-95th percentiles of the posterior distribution.}. While we find that the uncertainty in the angle between the kick and the line-of-sight $\alpha$ is fairly constant with $\alpha_{inj}$, the uncertainty in $\cos\alpha$ increases as $\alpha_{inj}$ does, leading to larger uncertainties for low values of $K_{r,inj}$, for which $|\cos \alpha|$ is small. 
Also, for fixed intrinsic parameters, the distribution of $\alpha$ is proportional to $\sin \alpha$, causing the posterior distributions to show larger tails toward the low $|K_{r,est}|$ end when $|K_{r,inj}|$ is large. This can make the peak of the distribution differ from the injected value in some cases. Similarly, the larger density of our bank in the low $q$ region (for which $|K|$ is low) can cause a selection bias toward $K_{r,est}=0$, resulting in the bimodal shape of some of the distributions. The peak near zero disappears as we raise the SNR and the poorly matching templates contributing to it are down-weighted. When $|K_{r,inj}| \geq 120km/s$ we can obtain non-biased estimates that exclude $K_{r}=0$ from $90\%$ credible interval of our posteriors.\\ 

\begin{figure}
\includegraphics[width=0.50\textwidth]{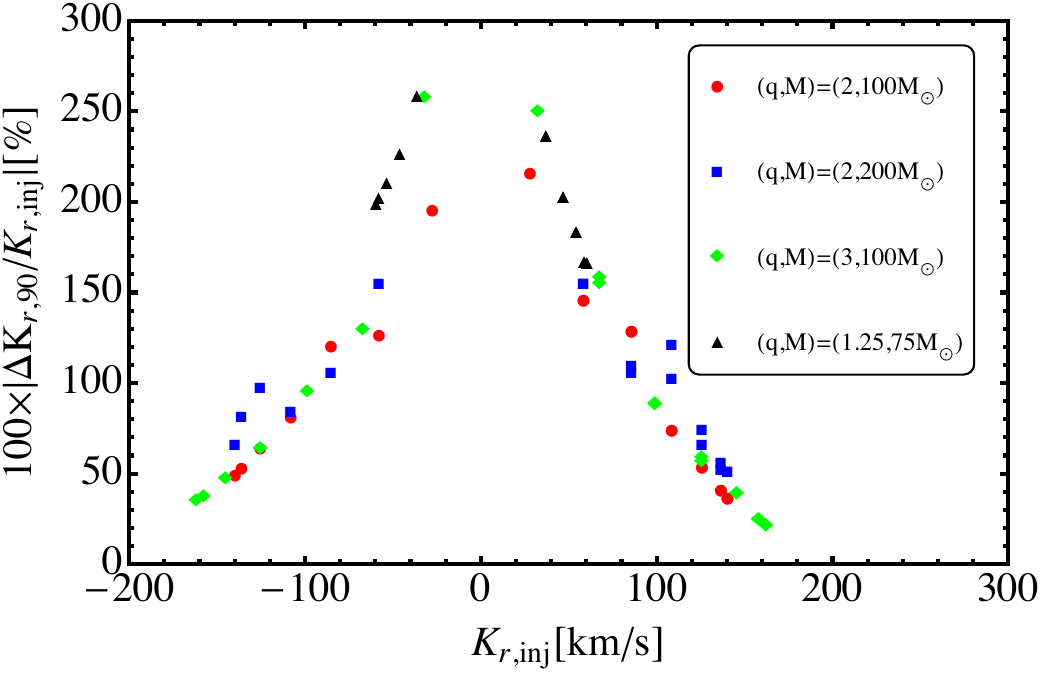}
\caption{\textbf{Uncertainty in the measurement of $K_r$}: Percent relative width of the $90\%$ credible intervals, $\Delta K_{r,90}$, of the posterior distributions for $K_r$ for signals emitted by various sources as a function of $K_r$. 
}
\label{fig:errors_summary}
\end{figure}

Fig.\ref{fig:errors_summary} shows the relative percent width of the $90\%$ credible intervals $\Delta K_{r,90}$, obtained for four different binaries with varying orientation as a function of $K_{r,inj}$. An accurate estimation of $K_r$ requires  the presence of a merger-ringdown signal with strong HMs to measure the orientation of the source (hence $\alpha$) and some inspiral cycles to infer the parameters of the binary (hence $|K|$) \cite{London:2017bcn,Graff:2015bba}. This happens for our $M=100_\odot$ cases (red and green), for which $K_r$ can be estimated better than $\Delta K_{r,90} = 60$km/s if $K_{r,inj} \geq 120$km/s. Instead, for the $200M_\odot$ case (blue), only a short ringdown portion of the signal is in band, which difficults to measure its intrinsic parameters and leads to slightly less accurate estimates of $K_r$. In any case, for all these sources we find orientations for which $K_r=0$ can be ruled out.
This is not true for the GW150914-like source, for which our measurement accuracy is never better than a $150\%$ of the injected value due to its weak HMs. In the best case, for $\alpha=0$, we obtain $K_{r,est}=32^{+ 8}_{-48}$km/s.
\\

\paragraph*{\textbf{Conclusions}} BBH coalescences radiate anisotropic GWs, imparting a recoil velocity, or kick, to the remnant BH. Detection of the recoil would provide the first observational evidence of net transport of linear momentum by gravitational waves away from their source. In this letter, we have prescribed and demonstrated a method to infer the component of the final BH kick along the line-of-sight. The method relies on exploiting the HMs of the GW emission to estimate the inclination and azimuthal angles of the binary. While the former has been commonly reported for BBH observations, the azimuth has been treated as a nuisance parameter due to its lack of clear physical meaning.  Expressing GW templates in their \textit{kick frame}, we redefine the azimuth as the angle between the final kick and the projection of the line-of-sight onto the orbital plane. Performing parameter estimation on aligned-spin sources using NR  templates, we have shown that for suitable BBHs, and at a SNR $\rho = 15$, modest kicks of $K_r \sim 120km/s$ can be estimated with $90\%$ credible intervals of around $60km/s$. This allows to rule out $K_r=0$, using a single Advanced LIGO detector working at its early sensitivity. Ruling out a zero kick requires the observation of highly inclined, unequal mass binaries with fairly large total mass. However, no such source has yet been observed \cite{Abbott:2017iws,CalderonBustillo:2017skv,PhysRevX.6.041015,TheLIGOScientific:2016wfe,Abbott:2016izl,Abbott:2016apu,Abbott:2016nmj,PhysRevLett.118.221101,Abbott:2017oio,Abbott:2017gyy} and it is unlikely that a zero kick can be ruled out using current BBH observations. With the Advanced LIGO and Virgo network about to commence its third observation run with improved sensitivity \cite{Aasi:2013wya} and the development of searches targeting these sources \cite{Harry:2017weg}, there is hope for such observations to happen in the near future.  Finally, our study is limited by the discreteness of our NR template bank, and should be taken as a proof-of-concept. Future application of these methods to BBH observations should implement continuous waveform families including the effects of spins and HMs \cite{Blackman:2017pcm,London:2017bcn,Cotesta:2018fcv}, after expressing waveforms in their kick frame.\\

\paragraph*{\textbf{Acknowledgements}}
It is our pleasure to thank Xisco Jimenez Forteza, Sebastian Khan and Tamara Bogdanovic for useful discussions. Also, we thank Katerina Chatziioannou and Ricardo Martinez-Garcia for comments on the manuscript.
JCB, JAC, PL and DS gratefully  acknowledge support from the NSF grants 1505824, 1505524, 1550461, XSEDE  TG-PHY120016. This research was also supported in part through research cyberinfrastructure resources and services provided by the Partnership for an Advanced Computing Environment (PACE) at the Georgia Institute of Technology~\cite{PACE}. This document has LIGO DCC number P1800179.

\bibliography{IMBBH.bib}

\end{document}